\documentclass[a4paper,11pt, onesided]{article}

\usepackage[ansinew]{inputenc}

\usepackage{amsfonts}
\usepackage{amsmath}
\usepackage{amsthm}

\usepackage{mathrsfs}
\usepackage{amssymb}

\def\PPh#1{\setbox0\hbox{$#1\rm
I$}\mathord{\vcenter{\ialign{$#1\rm##$\cr
I\cr\noalign{\nointerlineskip \vskip-0.541\ht0}P\cr}}}}

\def\Ph{{\mathpalette\PPh{}}}
\def\Ed{{\mathord{\mkern5mu\mathaccent"7020{\mkern-5mu\partial}}}}

\newtheorem{theorem}{Theorem}

\begin{document}

\title{The Chevreton Tensor and Einstein-Maxwell Spacetimes Conformal to Einstein Spaces}
\author{Göran Bergqvist and Ingemar Eriksson\\
       Matematiska institutionen, Linköpings universitet, \\
       SE-581 83 Linköping, Sweden \\
       gober@mai.liu.se, ineri@mai.liu.se}
\date{March 13, 2007}

\maketitle

\begin{abstract}
    In this paper we characterize the source-free Einstein-Maxwell spacetimes which have a trace-free Chevreton tensor.
    We show that this is equivalent to the Chevreton tensor being of pure-radiation type and that it restricts the
    spacetimes to Petrov types \textbf{N} or \textbf{O}. We prove that the trace of the Chevreton tensor is related to
    the Bach tensor and use this to find all Einstein-Maxwell spacetimes with a zero cosmological constant that have a
    vanishing Bach tensor. Among these spacetimes we then look for those which are conformal to Einstein spaces.
    We find that the electromagnetic field and the Weyl tensor must be aligned, and in the case that the electromagnetic
    field is null, the spacetime must be conformally Ricci-flat and all such solutions are known. In the non-null case,
    since the general solution is not known on closed form, we settle with giving the integrability conditions in the
    general case, but we do give new explicit examples of Einstein-Maxwell spacetimes that are conformal to Einstein
    spaces, and we also find examples where the vanishing of the Bach tensor does not imply that the spacetime is
    conformal to a $C$-space. The non-aligned Einstein-Maxwell spacetimes with vanishing Bach tensor are conformally
    $C$-spaces, but none of them are conformal to Einstein spaces.
\end{abstract}


\section{Introduction}
In this paper we first continue the characterization of the Chevreton tensor in four dimensional source-free
Einstein-Maxwell theory. This leads to a new relation between the Chevreton tensor and the Bach tensor, and this then
allows us to give a ``factorization'' of the Bach tensor, which we apply to investigate source-free
Einstein-Maxwell spacetimes that are conformal to Einstein spaces.

The Chevreton tensor was introduced in 1964 \cite{Chevreton1964} as an electromagnetic counterpart to the Bel-Robinson
superenergy tensor for the gravitational field \cite{Bel1958, Bel1959}. It is given by
\begin{equation}
    H_{abcd} \{ \nabla _{[1]} F_{[2]} \} = \frac{1}{2} ( T_{abcd} \{ \nabla _{[1]} F_{[2]} \} + T_{cdab} \{ \nabla _{[1]} F_{[2]} \} ),
\end{equation}
where
\begin{align}
  T_{abcd}\{ \nabla _{[1]} F_{[2]} \} =& -\nabla_{a} F_{ce}\nabla_b F_{d}{}^e- \nabla_{a} F_{de}\nabla_b F_{c}{}^e
 + g_{ab}\nabla_{f} F_{ce}\nabla^f F_{d}{}^e \nonumber\\
 &+ \frac{1}{2}g_{cd}\nabla_a F_{ef}\nabla_{b} F^{ef}
- \frac{1}{4} g_{ab}g_{cd} \nabla_e F_{fh}\nabla^e F^{fh}
\end{align}
is the basic electromagnetic superenergy tensor \cite{Senovilla2000}. Like all superenergy tensors it satisfies the
Dominant Property ($H_{abcd}v_1^a v_2^b v_3^c v_4^d \geq 0$, for all future-directed causal vectors $v_i^a$)
 \cite{Bergqvist1999, Senovilla2000}.
For source-free electromagnetic fields the Chevreton tensor is
completely symmetric in four dimensions \cite{Bergqvist2003}. In Special Relativity it is divergence-free and in
Einstein-Maxwell theory it has been shown to give rise to conserved currents for hypersurface orthogonal Killing
vectors \cite{Eriksson2006}. The trace of the Chevreton tensor, $H_{ab} = H_{abc}{}^c$, is a symmetric, trace-free,
and divergence-free two-index tensor \cite{Bergqvist2003}. The spinor form of the Chevreton tensor is given by
\begin{align}
    H_{abcd} \{ \nabla _{[1]} F_{[2]} \}
        = & \nabla _{AB'} \varphi_{CD} \nabla _{A'B} \bar \varphi _{C'D'}  + \nabla _{BA'} \varphi _{CD} \nabla _{B'A} \bar \varphi _{C'D'}  \nonumber\\
          & +\nabla _{CD'} \varphi _{AB} \nabla _{C'D} \bar \varphi _{A'B'}  + \nabla _{DC'} \varphi _{AB} \nabla _{D'C} \bar \varphi _{A'B'},
\end{align}
and the spinor form of the trace is given by
\begin{align} \label{eq:ChevretonTrace}
    H_{ab} = H_{abc}{}^c = -2\nabla_{CC'} \varphi_{AB} \nabla^{CC'} \bar\varphi_{A'B'}.
\end{align}

In section \ref{VanishingTrace} we study the case when the Chevreton tensor is trace-free. We prove that this is
equivalent to the Chevreton tensor being of pure-radiation type and that it restricts the spacetime to Petrov type
\textbf{N} or \textbf{O}. We show that the cosmological constant must be zero and we get a number of restrictions given
in terms of the GHP spin-coefficient formalism. In section 5 we use this to find all such spacetimes.

In section \ref{BachSection} we give the relation between the trace of the Chevreton tensor and the Bach tensor. Hence,
the solutions in section 5 are also those source-free Einstein-Maxwell spacetimes with a zero cosmological constant
that have a vanishing Bach tensor.

The problem of finding spacetimes that are conformally related, $\hat g_{ab} = e^{2\Omega}g_{ab}$, to Einstein spaces,
satisfying
\begin{align}
    \hat R_{ab} - \frac{1}{4}\hat R \hat g_{ab} = 0,
\end{align}
is an old problem that dates back to Brinkmann's work in 1924 \cite{Brinkmann1924}. The vanishing of the Bach tensor is
a necessary condition for finding such a transformation \cite{Kozameh1985}. Several attempts have been made in the
literature to find conditions that are sufficient, and useful conditions have been found for all Petrov types excluding
type \textbf{N} \cite{Kozameh1985, Wunsch1990}. For the special case of spacetimes conformally related to empty (or
Ricci-flat) spaces, $\hat R_{ab}=0$, sufficient conditions have been given for the type {\bf N} case by Czapor,
McLenaghan, and Wünsch \cite{Czapor2002}, but the general case of conformally Einstein spaces has proven to be very
hard to analyze. Szekeres \cite{Szekeres1963} does give sufficient conditions, but they involve solvability of
differential equations. In the $n$-dimensional case conditions have been found in the generic case
\cite{Gover2006, Listing2001}.

In section \ref{ConformallyEinstein} we look at the integrability conditions in Einstein-Maxwell spacetimes for finding
a scalar that transforms the spacetime into an Einstein space.

In section \ref{Solutions} we find the spacetimes for which the Chevreton tensor is of pure-radiation type. These
correspond to all Einstein-Maxwell spacetimes with a zero cosmological constant that have a vanishing Bach tensor. We
divide them into three cases. In the case when the electromagnetic field and the Weyl tensor are non-aligned we find
that there are no conformally Einstein space solutions. In the case of an aligned null electromagnetic field all
solutions are conformally Ricci-flat and they were previously found by Van den Bergh \cite{Bergh1986}. The aligned
non-null case contains a subset of conformally Einstein solutions. We are not able to fully determine this set, but we
do give some explicit solutions.


\subsection{Conventions}
We assume that spacetime is a four dimensional Lorentzian manifold of signature $(+,-,-,-)$ and we follow the
conventions of \cite{Penrose1984} and \cite{Penrose1986}, though we set $8\pi G = 1$. The Einstein-Maxwell equations
are given by
\begin{align} \label{eq:Einstein}
    R_{ab} - \frac{1}{4}R g_{ab} = -T_{ab} = 2 F_{ac} F_b{}^c - \frac{1}{2}g_{ab} F_{cd}F^{cd},
\end{align}
where $R=4\lambda$ and $\lambda$ is the cosmological constant. The electromagnetic field $F_{ab} = -F_{ba}$ satisfies
the source-free Maxwell equations
\begin{align}
    \nabla^a F_{ab} &= 0, \\
    \nabla_{[a}F_{bc]} &= 0.
\end{align}
The spinor version of these equations is given by
\begin{align}
    T_{ab} &= 4\varphi_{AB}\bar\varphi_{A'B'} = 2\Phi_{ABA'B'}, \label{eq:EinsteinMaxwell} \\
    \Lambda &= \frac{1}{6}\lambda, \\
    \nabla^{AA'}\varphi_{AB} &= 0,
\end{align}
where $\varphi_{AB}$ is the Maxwell spinor and $\Phi_{ABA'B'}$ is the Ricci spinor. We note that the last equation can
be written as
\begin{align} \label{eq:Maxwell}
  \nabla_{CC'}\varphi_{AB} = \nabla_{C'(C}\varphi_{AB)}.
\end{align}
The Bianchi identity is given by
\begin{align} \label{eq:Bianchi}
  \nabla^{AA'}\Psi_{ABCD} = 2 \bar\varphi_{E'}{}^{A'}\nabla_B{}^{E'}\varphi_{CD},
\end{align}
where $\Psi_{ABCD}$ is the Weyl spinor.

In the calculations we will also make use of the Newman-Penrose (NP) and Geroch-Held-Penrose (GHP) spin-coefficient
formalisms \cite{Penrose1984}. We will mainly use the GHP formalism since it offers several advantages over NP for
algebraic computations.


\section{Vanishing trace of the Chevreton tensor} \label{VanishingTrace}
In this section we investigate the consequences of the Chevreton tensor being trace-free, $H_{abc}{}^c =
H_{ab} = 0$. This is equivalent to
\begin{align}
    \nabla_{CC'}\varphi_{AB} \nabla^{CC'} \bar\varphi_{A'B'} = 0.
\end{align}
The simplest solution to this equation is that the electromagnetic field is covariantly constant,
$\nabla_{CC'}\varphi_{AB} = 0$, which is satisfied if and only if the Chevreton tensor is zero, $H_{abcd} = 0$
\cite{Senovilla2000}. The only solution with this property is the Bertotti-Robinson solution \cite{Stephani2003},
which is given by
\begin{align} \label{Metric:BertottiRobinson}
    {\rm d}s^2 = k^2\left( \sinh^2x{\rm d}t^2 - {\rm d}\theta^2 - \sin^2\theta{\rm d}\phi^2 - {\rm d}x^2 \right),
\end{align}
and is of Petrov type {\bf O}. We will thus assume that the Chevreton tensor is not identically zero, then
$\nabla_{CC'}\varphi_{AB} \neq 0$, and we have an equation of the type
\begin{align}
    & S_{ABCC'}\bar S_{A'B'}{}^{CC'} =0,
\end{align}
with solution \cite{Penrose1984}
\begin{align}
    S_{ABCC'} = (o_C o_{C'} + i \beta_C \bar\beta_{C'})\eta_{AB}.
\end{align}
We want to show that $S_{ABCC'}$ is proportional to $o_A o_B o_C o_{C'}$. Assume that $o_A \beta^A \neq 0$ and let $(o^A, \iota^A)$
be a spin-basis, then the symmetric spinor $\eta_{AB}$ can be decomposed as
\begin{align}
    \eta_{AB} = Ao_A o_B + B o_A\iota_B + B \iota_A o_B + C\iota_A\iota_B.
\end{align}
By the source-free Maxwell equations (\ref{eq:Maxwell}), $S_{AC}{}^C{}_{C'} = 0$, so
\begin{align}
    (o^C o_{C'} + i\beta^C\bar\beta_{C'})( Ao_A o_C + B o_A\iota_C + B \iota_A o_C + C\iota_A\iota_C) = 0.
\end{align}
Expanding this gives
\begin{align} \label{Temp1}
    &-Bo_Ao_{C'} - C\iota_Ao_{C'} + i\bar\beta_{C'}\beta^C(Ao_A o_C + B o_A\iota_C + B \iota_A o_C + C\iota_A\iota_C   )=0.
\end{align}
Contracting with $\bar\beta^{C'}$ gives
\begin{align}
    \bar\beta^{C'}o_{C'}(Bo_A + C\iota_A)  =0.
\end{align}
Thus, either $o_A$ and $\beta_A$ are proportional, or $B$ and $C$ are zero. For the second case, equation (\ref{Temp1})
also gives us that $o_A$ and $\beta_A$ are proportional. Hence, we have that
\begin{align}
    S_{ABCC'} = o_C o_{C'}\eta_{AB}.
\end{align}
The source-free Maxwell equations (\ref{eq:Maxwell}) imply that
\begin{align}
    & o_{C'} o^A \eta_{AB} = o_{C'}o^B \eta_{AB} = 0,
\end{align}
so
\begin{align} \label{eq:Rewriting}
    \nabla_{CC'} \varphi_{AB} = S_{ABCC'} = \eta o_C o_{C'} o_A o_B,
\end{align}
where the scalar $\eta$ is assumed to be nonzero. We note that the spin-boost weight of $\eta$ is $\{-3,-1\}$. Also,
with $l^a = o^A o^{A'}$, if the Chevreton tensor is trace-free, then it will be given by $H_{abcd} = 4 \eta\bar\eta
l_a l_b l_c l_d$. On the other hand, if the Chevreton tensor is given by $H_{abcd} \propto l_a l_b l_c l_d$, where
$l^a$ is a null vector, then it is trace-free. Hence,
\begin{theorem}
    The Chevreton tensor is of pure-radiation type, $H_{abcd} \propto l_a l_b l_c l_d$, where $l^a$ is a null vector,
    if and only if it is trace-free, $H_{abc}{}^c = 0$.
\end{theorem}

We can now substitute with equation (\ref{eq:Rewriting}) whenever we have a derivative of the Maxwell spinor present.
We start by examining the expression
\begin{align}
    \nabla_{C(C'}\nabla^C{}_{D')}\varphi_{AB} =& \Box_{C'D'}\varphi_{AB} =  -\Phi_{C'D'A}{}^{E}\varphi_{EB}
       - \Phi_{C'D'B'}{}^{E}\varphi_{AE} \nonumber\\
    =&-2\bar\varphi_{C'D'}(\varphi_A{}^E\varphi_{EB} + \varphi_B{}^E\varphi_{AE}) =0,
\end{align}
where in the last step we used the Einstein-Maxwell equations (\ref{eq:EinsteinMaxwell}). But from (\ref{eq:Rewriting})
we also have that
\begin{align}
    2\nabla_{C(C'}\nabla^C{}_{D')}\varphi_{AB}
    =& \nabla_{CC'}\nabla^C{}_{D'}\varphi_{AB} + \nabla_{CD'}\nabla^C{}_{C'}\varphi_{AB} \nonumber\\
    =&\nabla_{CC'}(\eta o_{D'}o^C o_A o_B) + \nabla_{CD'}( \eta o_{C'}o^C o_A o_B) \nonumber\\
    =&o_A o_B \eta (o^C\nabla_{CC'}o_{D'} + o_{D'}\nabla_{CC'}o^C + o^C\nabla_{CD'}o_{C'} + o_{C'}\nabla_{CD'}o^C  ) \nonumber\\
    &+o_A o_B( o^C o_{C'}\nabla_{CD'}\eta+o^C o_{D'}\nabla_{CC'}\eta  ) \nonumber\\
    &+\eta o_{D'}o^C (o_A\nabla_{CC'} o_B + o_B \nabla_{CC'} o_A) \nonumber\\
    & +\eta o_{C'}o^C (o_A\nabla_{CD'} o_B + o_B \nabla_{CD'} o_A) \nonumber\\
    =& 0.
\end{align}
Contracting this with $o^B$ gives
\begin{align}
    &  o_{D'} o^C o^B \nabla_{CC'} o_B + o_{C'}  o^C o^B \nabla_{CD'} o_B = 0.
\end{align}
Now, contracting with $o^{C'}$ gives
\begin{align}
    o^B o^C o^{C'} \nabla_{CC'} o_B = o^B \Ph o_B = \kappa = 0,
\end{align}
and contraction by $\iota^{C'} \iota^{D'}$ gives
\begin{align}
    o^B o^C \iota^{C'} \nabla_{CC'} o_B = o^B \Ed o_B = \sigma = 0.
\end{align}
The vanishing of $\kappa$ and $\sigma$ implies that $l^a = o^A o^{A'}$ generates a shear-free geodetic null congruence
\cite{Penrose1986}. The Bianchi identity (\ref{eq:Bianchi}) is now given by
\begin{align}
    \nabla^{AA'}\Psi_{ABCD}  = 2\bar\varphi_{E'}{}^{A'} \eta o^{E'} o_B o_C o_D.
\end{align}
Thus one contraction of $\nabla^{AA'}\Psi_{ABCD}$ with $o^B$ gives zero, and this together with the fact that $l^a$
generates a shear-free geodetic null congruence gives us, via the generalized Goldberg-Sachs theorem
\cite{Penrose1986}, that $\Psi_{ABCD}$ has $o^A$ as a four-fold repeated principal null direction,
\begin{align}
    \Psi_{ABCD} = \Psi_4 o_A o_B o_C o_D.
\end{align}
The spacetime is thus of Petrov type \textbf{N} or \textbf{O}.  We state this as a theorem,
\begin{theorem} \label{Theorem:PetrovType}
    If the Chevreton tensor in source-free Einstein-Maxwell theory is trace-free, then the spacetime is of Petrov type
    \textbf{N} or \textbf{O} and the electromagnetic field satisfies
    \begin{align}
        \nabla_{CC'}\varphi_{AB} = \eta o_{C'}o_C o_A o_B,
    \end{align}
    where in Petrov type \textbf{N}, $o^A$ coincides with the principal null direction of the Weyl spinor.
\end{theorem}
If the spacetime is of Petrov type \textbf{O}, i.e., $\Psi_{ABCD}=0$, then the Bianchi identity gives us that
$\varphi_{AB}o^B = 0$. Hence, the electromagnetic field is null, with $\varphi_{AB} = \phi_2 o_A o_B$, and the
solutions are given by \cite{Stephani2003}
\begin{align} \label{Metric:TypeOnull}
    {\rm d}s^2 = -{\rm d}x^2 - {\rm d}y^2 +2{\rm d}u{\rm d}v + \kappa_o(x^2+y^2)f^2(u){\rm d}u^2 / 2.
\end{align}

To continue the characterization we decompose the derivative of the Maxwell spinor into its spin-basis components.
First,
\begin{subequations}
\begin{align}
  \Ph\varphi_{AB} &= o^Co^{C'}\nabla_{CC'}\varphi_{AB} = 0, \\
  \Ph'\varphi_{AB} &= \iota^C\iota^{C'}\nabla_{CC'}\varphi_{AB} = \eta o_A o_B, \\
  \Ed\varphi_{AB} &= o^C\iota^{C'}\nabla_{CC'}\varphi_{AB} = 0, \\
  \Ed'\varphi_{AB} &= \iota^Co^{C'}\nabla_{CC'}\varphi_{AB} = 0.
\end{align}
\end{subequations}
The Maxwell spinor expanded in the spin-basis is given by
\begin{align}
  \varphi_{AB} = \phi_2 o_A o_B - \phi_1 (o_A\iota_B + \iota_A o_B) + \phi_0 \iota_A \iota_B.
\end{align}
Thus, the first equation gives
\begin{align}
    &o_A o_B \Ph\phi_2 + \phi_2(o_A \Ph o_B + o_B \Ph o_A ) - (o_A\iota_B+\iota_A o_B)\Ph\phi_1  \nonumber\\
    &-\phi_1 (o_A \Ph \iota_B + \iota_B \Ph o_A + o_B \Ph \iota_A + \iota_A \Ph o_B   ) \nonumber\\
    &+\iota_A\iota_B \Ph\Phi_0 + \phi_0 (\iota_A \Ph \iota_B + \iota_B \Ph \iota_A) = 0.
\end{align}
Contracting this with $o^A o^B$, $o^A \iota^B$, and $\iota^A
\iota^B$ gives
\begin{subequations} \label{maxekv}
\begin{align}
    \Ph\phi_0 &= 0,    \\
    \Ph\phi_1 &= -\phi_0 \tau', \\
    \Ph\phi_2 &= -2\phi_1 \tau'.
\end{align}
Likewise the three other equations give
\begin{align}
    \Ph' \phi_0 &= - 2\phi_1\tau,               & \Ed\phi_0 &= 0,             & \Ed'\phi_0  &= -2\phi_1\rho,  \\
    \Ph' \phi_1 &= - \phi_0\kappa'- \phi_2\tau, & \Ed\phi_1 &= -\phi_0\rho',  & \Ed' \phi_1 &= -\phi_0\sigma'- \phi_2 \rho, \\
    \Ph' \phi_2 &= \eta - 2\phi_1\kappa',       & \Ed\phi_2 &=-2\phi_1\rho',  & \Ed'\phi_2  &= -2\phi_1\sigma'.
\end{align}
\end{subequations}
With these equations the only information that remains of the Bianchi identity is
\begin{subequations}\label{bianchi1}
\begin{align}
    \Ph\Psi_4 &= \rho \Psi_4 - 2\bar\phi_0 \eta, \\
    \Ed \Psi_4 &= \tau \Psi_4 - 2\bar\phi_1 \eta.
\end{align}
\end{subequations}

Knowing all the components of the derivatives of the electromagnetic field allows us to calculate all commutators for
$\phi_0$, $\phi_1$, and $\phi_2$ to extract more information. We start with applying
commutators to $\phi_0$, which is a $\{2,0\}$ weighted quantity,
\begin{align}
  \left(\Ph\Ph'-\Ph'\Ph\right)\phi_0 = (\bar\tau-\tau')\Ed\phi_0 + (\tau-\bar\tau')\Ed'\phi_0 - 2(-\tau\tau'+\Phi_{11}-\Lambda)\phi_0.
\end{align}
Using the above equations (\ref{maxekv}) and the GHP equations give
\begin{align}
    &-2\phi_1\left( (\tau-\bar\tau')\rho + \Phi_{01}\right) +2\tau\tau'\phi_0 \nonumber\\
    =& -2\rho\phi_1(\tau-\bar\tau') + 2\tau\tau'\phi_0 + 2\Phi_{11} \phi_0 - 2\Lambda\phi_0,
\end{align}
or
\begin{align}
    \Lambda\phi_0 = 0,
\end{align}
where we have used $\Phi_{11}\phi_0 = \Phi_{01}\phi_1$. Next,
\begin{align}
    \left( \Ph'\Ed'-\Ed'\Ph'\right) \phi_0 = \bar\rho'\Ed'\phi_0 +\sigma'\Ed\phi_0 -\bar\tau\Ph'\phi_0
      - \kappa'\Ph\phi_0+2(\rho\kappa'-\tau\sigma')\phi_0,
\end{align}
which reduces to
\begin{align}
    \Lambda\phi_1=0.
\end{align}
The other commutators on $\phi_0$ are identically satisfied or reproduce the same conditions as those above.

Continuing with $\phi_1$, which is a $\{0,0\}$ weighted quantity, we get from the commutator
\begin{align}
    \left( \Ph'\Ed'-\Ed'\Ph' \right ) \phi_1 = \bar\rho'\Ed'\phi_1 +\sigma'\Ed\phi_1 -\bar\tau\Ph'\phi_1 - \kappa'\Ph\phi_1
\end{align}
that
\begin{align}
    \rho\eta + \phi_0\Psi_4 - 2\Lambda\phi_2=0.
\end{align}
The other commutators on $\phi_1$ are identically satisfied or reproduce the same conditions as already found.

Last, we calculate the commutators for $\phi_2$, which is a $\{-2,0\}$ weighted quantity. From
\begin{align}
    \left( \Ph\Ph'-\Ph'\Ph \right ) \phi_2 = (\bar\tau-\tau')\Ed\phi_2 + (\tau-\bar\tau')\Ed'\phi_2 + 2(-\tau\tau'+\Phi_{11}-\Lambda)\phi_2
\end{align}
we get
\begin{align}
  \Ph\eta = -2\Lambda\phi_2.
\end{align}
Next,
\begin{align}
    \left( \Ph'\Ed'-\Ed'\Ph' \right) \phi_2 =
       \bar\rho'\Ed'\phi_2 +\sigma'\Ed\phi_2 -\bar\tau\Ph'\phi_2 - \kappa'\Ph\phi_2-2(\rho\kappa'-\tau\sigma')\phi_2
\end{align}
gives
\begin{align}
    \Ed'\eta = \bar\tau\eta-2\phi_1\Psi_4.
\end{align}
From
\begin{align}
    \left( \Ph'\Ed - \Ed\Ph' \right ) \phi_2 = \rho'\Ed\phi_2 + \bar\sigma'\Ed'\phi_2 - \tau\Ph'\phi_2
           - \bar\kappa'\Ph\phi_2 - 2(-\rho'\tau + \Phi_{12})\phi_2
\end{align}
we get
\begin{align}
    \Ed\eta = \tau\eta.
\end{align}
Finally, we have that
\begin{align}
    \left( \Ed\Ed' - \Ed'\Ed \right)\phi_2 = (\bar\rho'-\rho') \Ph\phi_2 + (\rho-\bar\rho)\Ph'\phi_2 - 2(\rho\rho'-\Phi_{11}-\Lambda)\phi_2,
\end{align}
which gives us that
\begin{align}
    (\rho-\bar\rho)\eta + 2\Lambda\phi_2 = 0.
\end{align}
The two other commutators for $\phi_2$ give no new information.

Collecting the results from the commutators we have the following conditions
\begin{subequations}\label{commutator1temp}
\begin{align}
                               \phi_0 \Lambda &= 0, \\
                               \phi_1 \Lambda &= 0, \\
    \rho\eta + \phi_0 \Psi_4 - 2\Lambda\phi_2 &= 0, \\
        (\rho-\bar\rho)\eta + 2\Lambda \phi_2 &= 0, \\
                                      \Ph\eta &=  - 2\Lambda\phi_2, \\
                                      \Ed\eta &= \tau\eta, \\
                                     \Ed'\eta &= \bar\tau \eta-2\phi_1\Psi_4.
\end{align}
\end{subequations}
If $\Lambda \neq 0$, then this implies that $\phi_0 = \phi_1 = 0$. The Maxwell equations (\ref{maxekv}) then imply that
$\rho=0$ and this in turn, by (\ref{commutator1temp}), implies that $\phi_2 = 0$. Thus, if $\Lambda$ is non-zero, then the
electromagnetic field vanishes. In the case of a covariantly constant electromagnetic field we have $\eta = 0$ and we
get directly that $\Lambda = 0$. Hence,
\begin{theorem}
    If the Chevreton tensor of a source-free Einstein-Maxwell spacetime is trace-free, the cosmological constant must
    be zero, $\lambda=0$.
\end{theorem}
Hence, we have that
\begin{subequations} \label{commutator1}
\begin{align}
    \rho\Phi + \phi_0 \Psi_4  &= 0, \\
                         \rho &= \bar\rho, \\
                      \Ph\eta &= 0,\\
                      \Ed\eta &= \tau\eta, \\
                     \Ed'\eta &= \bar\tau\eta-2\phi_1\Psi_4.
\end{align}
\end{subequations}
Thus, source-free Einstein-Maxwell spacetimes with non-vanishing trace-free Chevreton tensor satisfy (\ref{maxekv}),
(\ref{bianchi1}), and (\ref{commutator1}). If we look at spacetimes with a zero cosmological constant, then we will see
in the next section that these are also the conditions for source-free Einstein-Maxwell spacetimes to have a vanishing
Bach tensor.


\section{Relation to the Bach tensor} \label{BachSection}
In this section we examine the relation between the Bach tensor and the trace of the Chevreton tensor, $H_{ab}$.
We can rewrite the trace (\ref{eq:ChevretonTrace}) by the Leibniz rule as
\begin{align}
    H_{ab} = -2 \nabla_{CC'}\varphi_{AB}\nabla^{CC'}\bar\varphi_{A'B'}
        = -\Box(\varphi_{AB}\bar\varphi_{A'B'}) + \varphi_{AB}\Box\bar\varphi_{A'B'}  + \bar\varphi_{A'B'}\Box\varphi_{AB},
\end{align}
where $\Box = \nabla^a\nabla_a$. Using the Maxwell wave equation \cite{Penrose1986},
\begin{align}
    \Box\varphi_{AB} = 2\Psi_{ABCD}\varphi^{CD} - 8\Lambda \varphi_{AB}, \label{eq:MaxwellWave}
\end{align}
 this equals
\begin{align}
  -\Box(\varphi_{AB}\bar\varphi_{A'B'}) + 2\varphi_{AB}\bar\varphi^{C'D'}\bar\Psi_{A'B'C'D'}
  + 2\bar\varphi_{A'B'}\varphi^{CD}\Psi_{ABCD} - 16\Lambda \varphi_{AB}\bar\varphi_{A'B'},
\end{align}
or, rewritten in terms of the energy-momentum tensor and the cosmological constant,
\begin{align}
  H_{ab} = -\frac{1}{4}\Box T_{ab} + \frac{1}{2}C_{acbd}T^{cd} - \frac{2}{3} \lambda T_{ab}.
\end{align}
If we now use the Einstein equations (\ref{eq:Einstein}) we get
\begin{align}
  H_{ab} = \frac{1}{4} \Box R_{ab} -\frac{1}{2}C_{acbd}R^{cd} + \frac{2}{3}\lambda(R_{ab} - \frac{1}{4}R g_{ab}),
\end{align}
which is a purely geometric description of $H_{ab}$. This relation was derived in \cite{Edgar2004}. Now, since $H_{ab}$
is divergence-free, symmetric and trace-free, and since it is a combination of the Einstein tensor, a term that is
quadratic in the Riemann tensor, and a double derivative of the Riemann tensor, $H_{ab}$ should be a linear combination
of the completely known tensors with these properties \cite{Balfagon2000, Bergman2005, Collinson1962}.

We will show that $H_{ab}$ is related to the conformally well-behaved Bach tensor, which is defined by
\cite{Penrose1986}
\begin{align} \label{eq:Bach}
  B_{ab} = \nabla^c \nabla^d C_{acbd} - \frac{1}{2}R^{cd}C_{acbd}.
\end{align}
The first term of the right-hand side can be rewritten as
\begin{align}
  \nabla^c\nabla^d C_{acbd} =& \nabla^{CC'}\nabla^{DD'} \varepsilon_{A'C'}\varepsilon_{B'D'} \Psi_{ABCD} +cc
     = \nabla^C{}_{A'} \nabla^D{}_{B'} \Psi_{ABCD} + cc \nonumber\\
   =& \nabla^C{}_{A'}(2\bar\varphi_{B'E'} \nabla_A{}^{E'}\varphi_{BC})  +cc \nonumber\\
   =& 2 \nabla^C{}_{A'}\bar\varphi_{B'E'} \nabla_A{}^{E'}\varphi_{BC} +2 \bar\varphi_{B'E'} \nabla^C{}_{A'}\nabla_A{}^{E'}\varphi_{BC} + cc \nonumber\\
   =& H_{ab} + 2 \bar\varphi_{B'E'} \nabla^C{}_{A'} \nabla_C{}^{E'} \varphi_{AB} + cc,
\end{align}
where the Bianchi identity (\ref{eq:Bianchi}) was used in the second step and ``$cc$" denotes the complex conjugate.
Now, by use of the Maxwell wave equation (\ref{eq:MaxwellWave}), Einstien's equations (\ref{eq:Einstein}), and the fact
that the trace of a symmetric spinor index pair vanishes,
\begin{align}
  \nabla^C{}_{A'} \nabla_C{}^{E'}\varphi_{AB} = & -\Box_{A'}{}^{E'} \varphi_{AB} - \frac{1}{2} \varepsilon_{A'}{}^{E'}\Box \varphi_{AB} \nonumber\\
    =& \Phi_{A'}{}^{E'}{}_A{}^F \varphi_{FB} +\Phi_{A'}{}^{E'}{}_B{}^F \varphi_{AF} - \varepsilon_{A'}{}^{E'}\varphi^{CD}\Psi_{ABCD} + 4\Lambda
      \varepsilon_{A'}{}^{E'}\varphi_{AB} \nonumber\\
    =& - \varepsilon_{A'}{}^{E'}\varphi^{CD}\Psi_{ABCD} + 4\Lambda \varepsilon_{A'}{}^{E'}\varphi_{AB}.
\end{align}
Hence,
\begin{align}
  \nabla^c\nabla^d C_{acbd} =& H_{ab} - 2 \bar\varphi_{A'B'} \varphi^{CD} \Psi_{ABCD} + 8\Lambda \varphi_{AB}\bar\varphi_{A'B'}  +cc. \nonumber\\
  =& 2 H_{ab} + \frac{1}{2}R^{cd} C_{acbd} + 4 \Lambda T_{ab},
\end{align}
or, using (\ref{eq:Bach}),
\begin{align}
  B_{ab} = 2 H_{ab} + \frac{2}{3} \lambda T_{ab}.
\end{align}
Note that we have here the Bach tensor written completely in terms of the electromagnetic field and its covariant
derivative.

\begin{theorem} \label{BachTheorem}
In Einstein-Maxwell spacetimes with a source-free electromagnetic field, the Bach tensor, $B_{ab}$, is related to the
trace of the Chevreton tensor, $H_{ab}$, by
\begin{align}
  B_{ab} = 2 H_{ab} + \frac{2}{3} \lambda T_{ab},
\end{align}
where $T_{ab}$ is the energy-momentum tensor and $\lambda$ is the cosmological constant.
\end{theorem}

We note that this result is not related to the conformally invariant theory of gravitation and electromagnetism,
$B_{ab} = 16\pi\kappa^2 T_{ab}$, suggested by Merkulov \cite{Merkulov1984}, since we have used the Einstein-Maxwell
equations.


\section{Conformally Einstein spaces} \label{ConformallyEinstein}
In this section we look at the integrability conditions for a source-free Einstein-Maxwell spacetime to be conformal to
(i.e., have the same null cone structure as) an Einstein space.

A spacetime is said to be conformal to an Einstein space if there exists a conformal transformation
$\hat g_{ab} = e^{2\Omega} g_{ab}$ such that in the transformed spacetime we have
\begin{align}
    \hat R_{ab} -\frac{1}{4}\hat R \hat g_{ab} = 0.
\end{align}
The vanishing of the Bach tensor is a necessary - but not sufficient - condition for a spacetime to be conformally
related to an Einstein space \cite{Kozameh1985}. Useful sufficiency conditions have been given for Petrov types
\textbf{I}, \textbf{II}, \textbf{D}, where the invariant $J= \Psi_{ABCD}\Psi^{CD}{}_{EF}\Psi^{EFAB} \neq 0$ , and
\textbf{III}, where $J=0$ \cite{Kozameh1985,Wunsch1990}, but Petrov type {\bf N} (also $J=0$) has proven to be very difficult to
analyze \cite{Czapor2002,Szekeres1963}. All Petrov type \textbf{O} spacetimes are conformally flat \cite{Schouten1921}, so they
are trivially conformally Einstein spaces. If we restrict our study to Einstein-Maxwell spacetimes with a zero
cosmological constant, then theorem \ref{BachTheorem} tells us that the vanishing of the Bach tensor is equivalent to
the Chevreton tensor being trace-free. For the case of Einstein-Maxwell spacetimes with a cosmological constant, the
vanishing of the Bach tensor gives the equation
\begin{align}
    \nabla_{CC'} \varphi_{AB} \nabla^{CC'} \bar\varphi_{A'B'} = 4\Lambda \varphi_{AB}\bar\varphi_{A'B'},
\end{align}
which is a much more difficult equation to analyze. Hence, we assume that the cosmological
constant is zero and theorem \ref{Theorem:PetrovType} then restricts us to Petrov type \textbf{N}.

To find the conformally Einstein solutions we will use the fact that they are a subclass of the conformally $C$-spaces,
which satisfy
\begin{align}
    \hat \nabla^a \hat C_{abcd} = 0.
\end{align}
Hence, a spacetime is conformal to a C-space if and only if there exists a smooth real scalar $\Omega$ such that
\begin{align}
    \nabla^{AA'}(e^\Omega \Psi_{ABCD}) = 0
\end{align}
and this implies for Petrov type {\bf N} that \cite{Czapor2002}
\begin{align}
    \kappa &= \sigma = 0, \nonumber\\
    \Ph \Omega &= \rho - \frac{\Ph\Psi_4}{\Psi_4}, \nonumber\\
    \Ed \Omega &= \tau - \frac{\Ed\Psi_4}{\Psi_4}.
\end{align}
In our case the Bianchi equations (\ref{bianchi1}) together with the fact that $\Omega$ is real valued give us that
\begin{align}
    \Ph\Omega &= 2\frac{\bar\phi_0\eta}{\Psi_4} = 2\frac{\phi_0\bar\eta}{\bar\Psi_4}, \nonumber\\
    \Ed\Omega &= 2\frac{\bar\phi_1\eta}{\Psi_4}.
\end{align}
By equations (\ref{commutator1}) the above reality condition is already satisfied. Let $\Ph'\Omega = A$, which is of
spin-boost weight $\{-1,-1\}$. We now use the commutators to extract integrability conditions for a conformally
$C$-space. The first commutator for $\Omega$ gives us
\begin{align}
    (\Ph\Ph' - \Ph'\Ph)\Omega = (\bar\tau - \tau')\Ed\Omega + (\tau - \bar\tau')\Ed'\Omega,
\end{align}
or
\begin{align}
    \Ph A  = 2\Ph'(\frac{\bar\phi_0\eta}{\Psi_4}) + 2(\bar\tau - \tau')\frac{\bar\phi_1\eta}{\Psi_4}
                                                  + 2(\tau - \bar\tau')\frac{\phi_1\bar\eta}{\bar\Psi_4}.
\end{align}
The second commutator
\begin{align}
    (\Ph\Ed - \Ed\Ph)\Omega = \rho\Ed\Omega - \bar\tau'\Ph\Omega
\end{align}
is identically satisfied. The third commutator
\begin{align}
    (\Ph'\Ed' - \Ed'\Ph')\Omega = \bar\rho'\Ed'\Omega + \sigma'\Ed\Omega - \bar\tau\Ph'\Omega - \kappa'\Ph\Omega
\end{align}
gives us that
\begin{align}
  \Ed'A = 2\Ph'(\frac{\phi_1\bar\eta}{\bar\Psi_4}) - 2\frac{\bar\rho'\phi_1\bar\eta}{\bar\Psi_4} - 2\frac{\sigma'\bar\phi_1\eta}{\Psi_4}
         + 2\frac{\kappa'\bar\phi_0\eta}{\Psi_4} +\bar\tau A.
\end{align}
The last commutator
\begin{align}
    (\Ed\Ed'-\Ed'\Ed)\Omega = (\bar\rho' - \rho')\Ph\Omega
\end{align}
gives us that
\begin{align}
  \frac{\bar\phi_1\eta}{\Psi_4^2}\Ed'\Psi_4 - \frac{\bar\tau\bar\phi_1\eta}{\Psi_4} =
     \frac{\phi_1\bar\eta}{\bar\Psi_4^2}\Ed\bar\Psi_4 - \frac{\tau\phi_1\bar\eta}{\bar\Psi_4}
\end{align}
If we further demand that the spacetime is conformal to an Einstein space then \cite{Penrose1984}
\begin{align}
    \Phi_{ABA'B'} = \nabla_{A(A'}\Upsilon_{B')B} - \Upsilon_{A(A'}\Upsilon_{B')B},
\end{align}
where $\Upsilon_{AA'} = \nabla_a\Omega$, must be satisfied. The $o^Ao^B o^{A'}o^{B'}$, $o^A o^B \iota^{A'} o^{B'}$, and
$o^A o^B \iota^{A'} \iota^{B'}$ components are identically satisfied, while $o^A \iota^B \iota^{A'} \iota^{B'}$ gives
\begin{align}
    \Ed A = 2\frac{\bar\phi_1\eta A}{\Psi_4} + 2\phi_1\bar\phi_2 - 2\frac{\bar\sigma'\phi_1\bar\eta}{\bar\Psi_4}
           -2\frac{\rho'\bar\phi_1\eta}{\Psi_4},
\end{align}
$\iota^A\iota^B \iota^{A'}\iota^{B'}$ gives
\begin{align}
    \Ph' A = 2\phi_2\bar\phi_2 + A^2 - 2\frac{\bar\kappa'\phi_1\bar\eta}{\bar\Psi_4} - 2\frac{\kappa'\bar\phi_1\eta}{\Psi_4},
\end{align}
and finally $o^A\iota^B o^{A'} \iota^{B'}$ gives
\begin{align}
    \Ph A = 4\phi_1\bar\phi_1 - 2\frac{\tau'\bar\phi_1\eta}{\Psi_4} - 2\frac{\bar\tau'\phi_1\bar\eta}{\bar\Psi_4}
            -2\Ed(\frac{\phi_1\bar\eta}{\bar\Psi_4}) - \rho A - 2\frac{\rho'\bar\phi_0\eta}{\Psi_4} +
             2\frac{A\bar\phi_0\eta}{\Psi_4} +  4\frac{\phi_1\bar\phi_1\eta\bar\eta}{\Psi_4\bar\Psi_4}.
\end{align}
If the space is also conformally Ricci-flat, $\hat R_{ab}=0$, then
\begin{align}\label{eq:Ricciflat}
    \nabla^a\nabla_a \Omega + \nabla^a\Omega \nabla_a\Omega = 0.
\end{align}


\section{Solutions} \label{Solutions}
In this section we will give the source-free Einstein-Maxwell spacetimes with trace-free Chevreton tensor. These are
spacetimes with a zero cosmological constant and vanishing Bach tensor. Among these we will then look for conformally
Einstein space solutions.

First, the two Petrov type {\bf O} solutions (\ref{Metric:BertottiRobinson}) and (\ref{Metric:TypeOnull}) are both
conformally flat, $\hat R_{abcd} = 0$ and are thus trivially conformally Einstein spaces.

The Petrov type {\bf N} solutions are divided into different cases, depending on whether the electromagnetic field and
the Weyl tensor are aligned or not. They are said to be aligned if \cite{Stephani2003}
\begin{align}
  \Psi_{ABCD}\varphi^{AB} = 0,
\end{align}
or
\begin{align}
  o_A o_B \varphi^{AB} = 0.
\end{align}
This means that the Weyl tensor and the electromagnetic field have a common principal null direction, $o_A$. The aligned case
will be further divided into a null or non-null electromagnetic field.


\subsection{Non-aligned}

The general Petrov type {\bf N} non-aligned non-null Einstein-Maxwell solution \cite{Stephani2003} was found by Cahen
and Leroy \cite{Cahen1966} and by Szekeres \cite{Szekeres1966}, and it is given by
\begin{align} \label{Metric:Non-aligned}
 {\rm d}s^2 =& -\frac{1}{2}\cos^2(ar)\left( {\rm d}x^2 + {\rm d}y^2 \right ) +2b\left(2r+\frac{\sin(2ar)}{a}\right){\rm d}u{\rm d}x \nonumber\\
  &+4{\rm d}u{\rm d}r - \frac{4}{a^2}\left( 2b^2\sin^2(ar)-2{\rm e}^{2u}-raa'_u\right){\rm d}u^2,
\end{align}
where
\begin{align}
  a=a(x,u) &= g(u){\rm cosech}\left({\rm e}^u x + f(u)\right), \\
  b=b(x,u) &= -{\rm e}^u \coth\left({\rm e}^u x + f(u)\right),
\end{align}
and $g=g(u) \neq 0$ and $f=f(u)$ are arbitrary \footnote{The coordinate
transformation $x' = x + e^{-u}f(u)$ makes computations in GRTensorII considerably faster.}.
As can be checked, with for example GRTensorII
\footnote{This is a package which runs within MAPLE and is available from address http://grtensor.org.}
, \emph{the Bach tensor
vanishes for these spacetimes}. This result does not seem to have been noticed before.

We will now analyze this case for conformal Einstein space solutions. We have that $\phi_0 \neq 0$ and by a null
rotation about $o_A$ we set $\phi_1 = 0$. From equation (\ref{commutator1}) we have that
\begin{align}
    \eta = -\frac{\phi_0 \Psi_4}{\rho},
\end{align}
and taking a $\Ed'$-derivative and using (\ref{commutator1}) we get
\begin{align}
  \Ed'\Psi_4 = \bar\tau\Psi_4.
\end{align}
Our Maxwell equations (\ref{maxekv}) give us
\begin{align}
   \Ph \phi_0 &= 0,   &   \Ph \phi_2 &= 0, \nonumber\\
   \Ed \phi_0 &= 0,   &   \Ed \phi_2 &= 0, \nonumber\\
   \Ed'\phi_0 &= 0,   &   \Ed'\phi_2 &= 0, \nonumber\\
   \Ph'\phi_0 &= 0,   &   \Ph'\phi_2 &= -\frac{\phi_0\Psi_4}{\rho},
\end{align}
and
\begin{align}
  \phi_0\kappa' &= -\phi_2\tau,   &      \phi_0\sigma' &= -\phi_2\rho,  &   \rho' = \tau' &= 0.
\end{align}
The equations for the conformal scalar are
\begin{align}
  \Ph \Omega &= -\frac{\Phi_{00}}{\rho}, \\
  \Ed \Omega &= 0, \\
  \Ph'\Omega &= A.
\end{align}
The integrability conditions for conformal C-space are
\begin{align}
  \Ph A &= \frac{\Phi_{00}}{\rho^2}\Ph'\rho, \label{temp1}\\
  \Ed'A &= \frac{\phi_2\tau\Phi_{00}}{\phi_0\rho} + \bar\tau A,
\end{align}
while the conditions for conformally Einstein space give us that
\begin{align}
  \Ph A &= -\rho A - \frac{A\Phi_{00}}{\rho}, \label{temp2} \\
  \Ed A &= 0, \\
  \Ph'A &= \Phi_{22} + A^2. \label{temp7}
\end{align}
Hence,
\begin{align}
  A        &= - \frac{\phi_2\tau\Phi_{00}}{\phi_0\rho\bar\tau},
\end{align}
and we thus get the following conditions
\begin{align}
      \Ph'\rho &= \frac{\phi_2 \tau}{\phi_0 \bar\tau}(\rho^2 + \Phi_{00}), \\
        \Psi_4 &= \frac{\phi_2\rho}{\phi_0} \left( \frac{\Ph'\tau}{\tau} -\frac{\Ph'\bar\tau}{\bar\tau} \right), \label{brokencondition}\\
   \frac{\phi_2\tau}{\phi_0\bar\tau} &= \frac{\bar\phi_2\bar\tau}{\bar\phi_0\tau}.
\end{align}
Given these conditions the commutators for $A$ and $\rho$ are identically satisfied. We see that the gradient of $\Omega$
is proportional to the gradient of $\rho$,
\begin{align}
  \Ph \Omega &= - \frac{\Phi_{00}}{\rho(\rho^2 + \Phi_{00})} \Ph \rho, \nonumber\\
  \Ed \Omega &= - \frac{\Phi_{00}}{\rho(\rho^2 + \Phi_{00})} \Ed \rho, \nonumber\\
  \Ed'\Omega &= - \frac{\Phi_{00}}{\rho(\rho^2 + \Phi_{00})} \Ed'\rho, \nonumber\\
  \Ph'\Omega &= - \frac{\Phi_{00}}{\rho(\rho^2 + \Phi_{00})} \Ph'\rho,
\end{align}
and since $\Phi_{00}$ is constant with respect to the GHP derivatives, we can integrate this to
\begin{align}
  \Omega = \int{\frac{\Phi_{00}}{\rho(\rho^2 + \Phi_{00})}d\rho } = \frac{1}{2} \ln \left( \frac{\rho^2 + \Phi_{00}}{\rho^2} \right ) + C,
\end{align}
where the constant $C$ can be set to zero.

The null rotated ($\phi_1 = 0$) and coordinate transformed form of the metric (\ref{Metric:Non-aligned}) is given by
\begin{align}
    {\rm d}s^2 =& -\frac{1}{2}\cos^2(ar)\left({\rm d}x^2 +{\rm d}y^2\right)
                    +\left(  4 r b + 2\frac{ b \sin(2ar)}{a} + \cos^2(ar)h \right){\rm d}u{\rm d}x +4 {\rm d}u{\rm d}r \nonumber\\
                & -\frac{1}{2}\left( \cos^2(ar)h^2 + 4\frac{bh\sin(2ar)}{a} - 16\frac{b^2\cos^2(ar)}{a^2} + 16\frac{e^{2u}}{g^2}
                     -8r\frac{g'_u}{g} - 8rbx           \right){\rm d}u^2,
\end{align}
where
\begin{align}
    a &= a(x,u) = g(u){\rm cosech}(e^u x), \\
    b &= b(x,u) = -e^u\coth(e^u x),
\end{align}
and $g=g(u) \neq 0$ and $h=h(u)$ are arbitrary. For this metric, the conformal scalar $\Omega$ would be given by
\begin{align}
    \Omega = -\frac{1}{2} \ln  \sin^2(ar),
\end{align}
but the conformal metric $\hat g_{ab} = e^{2\Omega}g_{ab}$ has $\hat\Phi_{22}\neq 0$ and is thus not an Einstein space.
This is due to the fact that among the integrability conditions given above, (\ref{brokencondition}) is not satisfied.
We note however that this metric represents a conformally $C$-space with vanishing Bach tensor and we have a situation
which differs from Petrov types \textbf{I}, \textbf{II}, and \textbf{D}. If these spacetimes are conformal to
$C$-spaces and have a vanishing Bach tensor then they are conformal to Einstein spaces \cite{Kozameh1985}.

Given that the metric (\ref{Metric:Non-aligned}) represents the general non-aligned solution, we draw the conclusion
that there are no Einstein-Maxwell spacetimes conformal to Einstein spaces in this class. We were unable to find a
contradiction directly by using the commutators, though we note that if we restrict to conformally Ricci-flat spaces
then (\ref{eq:Ricciflat}) yields
\begin{align}
    A( \rho^2 + \Phi_{00} ) = 0,
\end{align}
which is a contradiction.


\subsection{Aligned, non-null electromagnetic field}
Here, $\phi_0 = 0$, $\phi_1 \neq 0$. By a null rotation we set $\phi_2 = 0$ and the Maxwell equations (\ref{maxekv})
are reduced to
\begin{align}
    \rho' = \tau' &= \sigma' = 0, \\
    \eta &= 2\kappa'\phi_1.
\end{align}
The general aligned non-null solutions are known \cite{Cahen1966, Szekeres1966}, though not given on closed form. The
solutions are quite complicated, so we therefore integrated the subset with trace-free Chevreton tensor directly. The
integration procedure is given in appendix \ref{solutionappendix}, and we find that our solutions are given (with
$x^3=x$ and $x^4 = y$) by
\begin{align} \label{metric:Alignednonnull}
    {\rm d}s^2 = 2(\Phi_{11}r^2-U^\circ(x,y,u)){\rm d}u^2 + 2{\rm d}u{\rm d}r -\frac{1}{2P(x,y)^2}( {\rm d}x^2 + {\rm d}y^2),
\end{align}
with
\begin{align}
    P^2 \left( \frac{\partial^2}{\partial x^2} + \frac{\partial^2}{\partial y^2} \right ) \ln P &= \Phi_{11}, \label{eq:Pcondition2}   \\
    \left( \frac{\partial^2}{\partial x^2} + \frac{\partial^2}{\partial y^2} \right ) U^\circ &= 0.
\end{align}
This metric was given in \cite{Szekeres1966} (without stating that $P'_u = 0$), though it was not known that the Bach
tensor vanished. If we look for conformal Einstein space solutions, the conformal scalar should satisfy
\begin{align}
    \Ph  \Omega &= 0, \\
    \Ed  \Omega &= 2\Phi_{11}\frac{\kappa'}{\Psi_4}, \\
    \Ph' \Omega &= A.
\end{align}
The integrability conditions for conformally $C$-spaces are
\begin{align}
    \Ph A &= 0, \\
    \Ed A &= 2\Phi_{11} \Ph'( \frac{\kappa'}{\Psi_4} ), \\
    \frac{\kappa'}{\Psi_4^2}\Ed'\Psi_4 &= \frac{ \bar\kappa'}{\bar\Psi_4^2}\Ed\bar\Psi_4.
\end{align}
The last condition here is not generally satisfied, as can be checked with for example GRtensorII. Hence, in the
general case, these spacetimes have a vanishing Bach tensor but are not conformally $C$-spaces and therefore
not conformally Einstein spaces.

The conditions for conformal Einstein spaces are
\begin{align}
    \Ed A &= \frac{2\Phi_{11}\kappa' A}{\Psi_4}, \\
    \Ph'A &= A^2 - \frac{2\bar\kappa'^2\Phi_{11}}{\bar\Psi_4} - \frac{2\kappa'^2\Phi_{11}}{\Psi_4}, \\
    \Ph A &= 2\Phi_{11} - 2\Phi_{11} \Ed( \frac{\bar\kappa'}{\bar\Psi_4}) + 4\Phi^2_{11}\frac{\kappa'\bar\kappa'}{\Psi_4\bar\Psi_4}.
\end{align}
The commutator $[\Ph,\Ph']A$ yields $2\Phi_{11}A = 0$, so we must take $A=0$, which reduces our
integrability conditions to
\begin{align}
    \Psi_4 \Ph' \kappa' &= \kappa' \Ph' \Psi_4, \label{eq:AlignedIntCond1} \\
    \Ed' \Psi_4 &= -2\Psi_4 \left( \frac{\Psi_4}{\kappa'} + \frac{\bar\kappa'\Phi_{11}}{\bar\Psi_4} \right ), \label{eq:AlignedIntCond2}\\
    \Psi_4 \bar\kappa'^2 + \bar\Psi_4\kappa'^2 &= 0.  \label{eq:AlignedIntCond3}
\end{align}
The equation for conformally Ricci-flat space yields
\begin{align}
    \Psi_4\bar\Psi_4 + 4\Phi_{11}\kappa'\bar\kappa' = 0,
\end{align}
so conformally Einstein spaces in this class must be proper conformally Einstein spaces, i.e., not conformally
Ricci-flat.

The first integrability condition (\ref{eq:AlignedIntCond1}) translates (using $z = x+iy$ and the results from
appendix \ref{solutionappendix}) into
\begin{align}
    \frac{\partial^2 U^\circ}{\partial z^2}\frac{\partial^2 U^\circ}{\partial z\partial u} =
            \frac{\partial U^\circ}{\partial z}\frac{\partial^3 U^\circ}{\partial z^2\partial u},
\end{align}
with solution
\begin{align}\label{eq:Usol}
    U^\circ = f(u)g(z) + \overline {f(u) g(z)}.
\end{align}
In the simple case when $g(z) = z$ the  third integrability condition translates to
\begin{align}
    \overline{ f(u)} \frac{\partial P}{\partial z} + f(u) \frac{\partial P}{\partial\bar z} = 0,
\end{align}
which gives
\begin{align}
    U^\circ &= h(u)( az +\bar a \bar z),   &    \bar a \frac{\partial P}{\partial z} + a \frac{\partial P}{\partial\bar z} = 0,
\end{align}
where $h(u)$ is real and $a$ is a constant satisfying $|a| =1$. We rotate the $xy$-plane such that
\begin{align}
    U^\circ = h(u)x. \label{metric:Alignednonnull_c1}
\end{align}
This then implies that $P = P(y)$ and we can give the explicit solution to equation (\ref{eq:Pcondition2}),
\begin{align}
    P = \frac{1}{2} \frac{\Phi_{11} + e^{2C_1 y + 2C_2}}{C_1 e^{C_1 y + C_2} }, \label{metric:Alignednonnull_c2}
\end{align}
where $C_1$ and $C_2$ are constants. The conformal scalar can then be integrated to
\begin{align}
    \Omega = \ln \left( \frac{\Phi_{11} + e^{2C_1 y + 2C_2}}{\Phi_{11} - e^{2C_1 y + 2C_2}}  \right),
\end{align}
and it transforms the spacetime into an Einstein space. Hence, the metric (\ref{metric:Alignednonnull}) satisfying
(\ref{metric:Alignednonnull_c1}) and (\ref{metric:Alignednonnull_c2}) is conformal to an Einstein space.

In the general case (\ref{eq:Usol}), we are not able to solve the differential equations of the integrability
conditions. The third integrability condition (\ref{eq:AlignedIntCond3}) becomes
\begin{align}
    \overline{ f(u)} \left(\frac{\partial \bar g}{\partial \bar z}\right)^2 \left( 2 \frac{\partial P}{\partial z}\frac{\partial g}{\partial z}
        + P\frac{\partial^2  g}{\partial  z^2}     \right ) +
         f(u)\left(\frac{\partial g}{\partial z}\right)^2 \left( 2 \frac{\partial P}{\partial \bar z}\frac{\partial \bar g}{\partial \bar z}
        + P\frac{\partial^2 \bar g}{\partial \bar z^2}     \right ) = 0,
\end{align}
which implies
\begin{align}
    U^\circ = h(u)( a g(z) + \overline {a g(z)} ),
\end{align}
where $h(u)$ is real and $a$ is a constant satisfying $|a| =1$. (The other possibility, that
$2 \frac{\partial P}{\partial z}\frac{\partial g}{\partial z} + P\frac{\partial^2  g}{\partial  z^2} = 0,$ yields
$\frac{\partial^2 \ln P}{\partial z \partial\bar z} = 0$, which is a contradiction.)

The second integrability condition (\ref{eq:AlignedIntCond3}) yields the equation
\begin{align}
    \frac{\partial g}{\partial z} \left(\frac{\partial P}{\partial z}\right)^2 + P\frac{\partial P}{\partial z}\frac{\partial^2 g}{\partial z^2}
    -P\frac{\partial^2 P}{\partial z^2}\frac{\partial g}{\partial z} - \frac{1}{2}P^2\frac{\partial^3 g}{\partial z^3}
    + \frac{P^2\left(\frac{\partial^2 g}{\partial z^2}\right)^2}{\frac{\partial g}{\partial z}}
    + \frac{1}{4}\Phi_{11}\frac{a\left(\frac{\partial g}{\partial z}\right)^2}{\bar a\frac{\partial \bar g}{\partial \bar z}} = 0,
\end{align}
which we are not able to analyze further. From the simple case $g(z) =z$ we know at least that the set of conformally
Einstein spaces in this class is not empty.


\subsection{Aligned, null electromagnetic field}
Here, $\phi_0 = \phi_1 = 0$ and our Maxwell equations (\ref{maxekv}) give us that $\rho = \tau = 0$. As we see from the
Bianchi identity (\ref{eq:Bianchi}), these spacetimes are $C$-spaces. The integrability conditions for the conformal
scalar are then trivially
\begin{align}
    \Ph \Omega &= 0,    &    \Ph A = 0, \\
    \Ed \Omega &= 0,    &    \Ed A = 0, \\
    \Ph'\Omega &= A.
\end{align}
The conditions for conformally Einstein spaces are
\begin{align}
    \Ph'A = \Phi_{22} + A^2.
\end{align}
The commutators for $A$ are identically satisfied, so we get no further integrability conditions. Hence, all these
spacetimes are conformally Einstein spaces. In fact, since the conformal scalar here satisfies $\nabla^a\nabla_a \Omega
= \nabla^a\Omega\nabla_a\Omega  = 0$, these spacetimes are conformally Ricci-flat (\ref{eq:Ricciflat}). These solutions
were found by Van den Bergh \cite{Bergh1986}. They belong to the family of \emph{pp} waves and are given by
\begin{align} \label{Metric:vandenbergh}
  {\rm d}s^2 = - 2{\rm d}\zeta {\rm d}\bar\zeta + 2{\rm d}u {\rm d}v + 2 \left (
                    f(\zeta,u) + \overline{ f(\zeta,u)} + \kappa_0 \zeta\bar\zeta (F(u))^2 \right) {\rm d}u^2,
\end{align}
where $f(\zeta,u)$ is analytic in $\zeta$ and both $f(\zeta,u)$ and $F(u)$ depend arbitrarily on $u$. The coordinate
transformation $\zeta' = \zeta \sqrt{F(u)}$ brings the metric into the form
\begin{align}
  {\rm d}s^2 &= -2F(u){\rm d}\zeta {\rm d}\bar\zeta
         -  F'_u(u) {\rm d}u \left(\zeta {\rm d}\bar\zeta + \bar\zeta {\rm d}\zeta
         \right)
         + 2 {\rm d}u {\rm d}v \nonumber\\
       &  + 2\left( f(\zeta,u) + \overline {f(\zeta,u)} +  \kappa_0
         \zeta\bar\zeta F(u) - \zeta\bar\zeta \frac{ (F'_u(u))^2}{4F(u)} \right) {\rm
         d}u^2,
\end{align}
where $\Phi_{22} = \kappa_0$ and we can find an explicit solution for the conformal scalar $\Omega$ which gives the
Ricci-flat metric,
\begin{align}
  \Omega(u) =  - \frac{1}{2}\ln \left( \sin^2(u\sqrt{\kappa_0} + C)\right).
\end{align}


\section{Conclusion}

We have given the characterization of all source-free Einstein-Maxwell spacetimes for which the Chevreton
electromagnetic superenergy tensor is trace-free. This is equivalent to the Chevreton tensor being of pure-radiation
type,
\begin{align}
    H_{abcd} \propto l_a l_b l_c l_d,
\end{align}
where $l_a$ is a null vector. This also implies that the spacetime must be of Petrov type \textbf{N} or \textbf{O} and
that the cosmological constant must be zero. The connection between the trace of the Chevreton tensor and the Bach
tensor is given by
\begin{align}
  B_{ab} = 2 H_{ab} + \frac{2}{3} \lambda T_{ab}.
\end{align}
We find all source-free Einstein-Maxwell spacetimes with a zero cosmological constant that have a vanishing Bach
tensor. We use this to study the spacetimes which are conformally related to Einstein spaces. The result is given in
the following table
\begin{center}
\begin{tabular}{|c|l|c|l|c|}
    \hline
    {\bf Petrov type} &  {\bf Electromagnetic field} &  {\bf Chevreton}            & {\bf Conformally Einstein}  &  {\bf Metric}  \\
    \hline
        \textbf{O}    &  non-null                    &  0                          & conformally flat            &   (\ref{Metric:BertottiRobinson})\\
    \hline
        \textbf{O}    &  null                        &  $\propto l_a l_b l_c l_d$  & conformally flat            &   (\ref{Metric:TypeOnull})\\
    \hline
        \textbf{N}    &  non-aligned, non-null       &  $\propto l_a l_b l_c l_d$  & no                          &   (\ref{Metric:Non-aligned}) \\
    \hline
        \textbf{N}    &  aligned, non-null           &  $\propto l_a l_b l_c l_d$  & some                        &   (\ref{metric:Alignednonnull}) \\
    \hline
        \textbf{N}    &  aligned, null               &  $\propto l_a l_b l_c l_d$  & conformally Ricci-flat      &   (\ref{Metric:vandenbergh}) \\
    \hline
\end{tabular}
\end{center}
All these metrics are previously known. The first two cases are trivially conformally Einstein spaces, since they are
of Petrov type \textbf{O}.

The third case is interesting, because it is a conformally $C$-space with a vanishing Bach tensor, but it is not
conformal to an Einstein space. It seems that the only previously found example of a spacetime (not necessarily
Einstein-Maxwell) with these properties is the one found by Nurowski and Pleba\'nski \cite{Nurowski2001},
\begin{align}
  {\rm d}s^2 = \mathrm{e}^{2\phi}\left [ -{\rm d}x^2-{\rm d}y^2 +\frac{2}{3}({\rm d}x+y^3{\rm d}u)(y{\rm d}r+\frac{11}{9}{\rm d}x-\frac{1}{9}y^3{\rm d}u)\right].
\end{align}
It belongs to the Fefferman class of metrics which is of Petrov type \textbf{N} and the principal null direction
of the Weyl tensor is geodetic, shear-free, twisting, and a conformal Killing vector field.

In the fourth case we have a situation where we have spacetimes with a vanishing Bach tensor that in the general case
are not even conformally $C$-spaces. We have not been able to determine the full subset of spacetimes that are
conformal to Einstein spaces, but we found some explicit examples. These are proper conformally Einstein spaces, i.e.,
not conformally Ricci-flat, and to our knowledge this is the first example of such Einstein-Maxwell spacetimes.

In the last case we find that all spacetimes are necessary conformally Ricci-flat. These were previously found by Van
den Bergh \cite{Bergh1986}.


\section*{Acknowledgments}

The authors wish to thank José Senovilla and Brian Edgar for valuable comments and discussions.


\appendix

\section{Integration of the aligned non-null solution} \label{solutionappendix}
The metric (\ref{metric:Alignednonnull}) representing the aligned non-null Einstein-Maxwell spacetime with vanishing
Bach tensor was given in \cite{Szekeres1966}, though without details. We need the expressions for the spin coefficients
to analyze possible conformally Einstein space solutions and we therefore give a brief review of the integration in the
NP formalism. We follow the coordinate choices of \cite{Newman1962} quite closely.

With $l^\mu$ equal to a gradient and $m^\mu$, $n^\mu$ parallelly propagated, both $\phi_1$ and $\phi_2$ are independent
of $r$, so with a null rotation we set $\phi_2 = 0$ and we have the following setup
\begin{align}
    l^\mu &= \delta_2^\mu, \\
    n^\mu &= \delta_1^\mu + U\delta_2^\mu + X^i\delta_i^\mu, \\
    m^\mu &= \omega\delta_2^\mu + \xi^i\delta_i^\mu,
\end{align}
where $i = 3,4$ and $x^1 = u$, $x^2=r$ and
\begin{align}
    \kappa = \sigma = \rho = \tau = \mu = \lambda = \pi = \varepsilon = \Lambda = \phi_0 = \phi_2 = 0, \qquad \alpha = -\bar\beta, \qquad
      \eta = -2\phi_1\nu.
\end{align}
Radial integration yields
\begin{align}
    \Psi_4   &= \Psi_4^\circ,                 &   \xi^i   &= \xi^{\circ i}, \\
    \alpha   &= \alpha^\circ,                 &   \omega  &= \omega^\circ, \\
    \gamma   &= \Phi_{11}r + i\gamma^\circ,   &   X^i     &= X^{\circ i}, \\
    \nu      &= \nu^\circ,                    &   U       &= -\Phi_{11}r^2 + U^\circ,
\end{align}
where we have set $\gamma^\circ$ real by a shift of the $r$-origin. From the two NP equations
\begin{align}
       -\Delta\bar\alpha - \delta\gamma    &= \bar\alpha(\bar\gamma-\gamma), \\
                                 \Delta\alpha     -\bar\delta\gamma &= \alpha(\bar\gamma-\gamma)   ,
\end{align}
we extract $\omega = 0$. By use of coordinate freedom and a spatial rotation we set
\begin{align}
    \xi^{\circ 3} = P, \qquad \xi^{\circ 4} = iP,
\end{align}
with $P$ real. Now,
\begin{align}
    \delta = 2P\frac{\partial}{\partial \bar z}, \qquad \bar\delta = 2P\frac{\partial}{\partial z},
\end{align}
with $z = x^3 + i x^4$. The transverse metric equation
\begin{align}
     \delta X^{\circ i} - \Delta \xi^{\circ i} &= -2i\gamma^\circ\xi^{\circ i}
\end{align}
gives $\delta(X^{\circ 3} + i X^{\circ 4}) = 0$ and we can use our coordinate freedom to set $X^{\circ 3} =  X^{\circ 4} = 0$.
Left from the above equation is now
\begin{align}
    \Delta P = 2i \gamma^\circ P \qquad \Leftrightarrow \qquad   \frac{\partial P}{\partial u} = 2i\gamma^\circ P \qquad \Leftrightarrow \qquad
        \gamma^\circ =  \frac{\partial  P}{\partial u} = 0,
\end{align}
since both $P$ and $\gamma^\circ$ are real. From the remaining NP and metric equations we now get
\begin{align}
    \alpha^\circ &= \frac{\partial P}{\partial z},    &     \nu^\circ &= -2P\frac{\partial U^\circ}{\partial z},    &
       \Psi_4^\circ &= -4 \frac{\partial}{\partial z}\left( P^2\frac{\partial U^\circ}{\partial z} \right ),
\end{align}
and the conditions
\begin{align}
    4P^2 \frac{\partial^2 \ln P}{\partial z \partial \bar z} &= \Phi_{11},  \label{eq:Pcondition}\\
      \frac{\partial^2 U^\circ}{\partial z \partial \bar z} &= 0.
\end{align}


\end{document}